\documentclass[12pt]{article}
\usepackage{graphics}
\usepackage[dvips]{graphicx}
\usepackage{mathrsfs}
\usepackage{amssymb}
\usepackage{verbatim}
\usepackage{float}
\usepackage{slashed}
\usepackage{amsmath}
\usepackage{bm}         

\newcommand{\be}{\begin{equation}}
\newcommand{\ee}{\end{equation}}
\newcommand{\bea}{\begin{eqnarray}}
\newcommand{\eea}{\end{eqnarray}}

\def\4vol{{\int d^4x \sqrt{-g}}}

\def\beq{\begin{equation}}
\def\eeq{\end{equation}}
\def\bea{\begin{eqnarray}}
\def\eea{\end{eqnarray}}
\def\bitem{\begin{itemize}}
\def\eitem{\end{itemize}}

\newcommand{\nc}{\newcommand}
\nc{\nt}{\tilde{N}}
\nc{\ra}{\rightarrow}
\nc{\lsim}{\begin{array}{c}\,\sim\vspace{-21pt}\\< \end{array}}
\nc{\gsim}{\begin{array}{c}\sim\vspace{-21pt}\\> \end{array}}
\nc{\tnt}{\tilde{N}}
\nc{\tst}{\tilde{t}}

\nc{\LL}{L}
\nc{\vv}{\tilde{v}}
\nc{\tbf}{ AM:\textbf}

\title{
\vspace*{5mm} \Large\textbf{LHC Discovery Potential for Non-Standard Higgs Bosons in the 
$\bm{3b}$ Channel}\\
\vspace*{1.0cm}
\author{\textbf{Marcela Carena$^{a,b,c}$, Stefania Gori$^{a,d}$, Aurelio Juste$^e$,}\\
\textbf{Arjun Menon$^f$, Carlos E.M. Wagner$^{a,c,d}$ and Lian-Tao Wang$^{a,c}$} \\
~\\
\small\emph{$^a$Enrico Fermi Institute, University of Chicago, Chicago, IL 60637} \\
\small\emph{$^b$Theoretical  Physics Department, Fermilab, Batavia, IL 60510}\\
\small\emph{$^c$Kavli Institute for Cosmological Physics, University of Chicago, Chicago, IL 60637}\\
\small\emph{$^d$HEP Division, Argonne National Laboratory, 9700 Cass Ave., Argonne, IL 60439}\\
\small\emph{$^e$Instituci\'{o} Catalana de Recerca i Estudis Avan\c{c}ats (ICREA) and}\\
\small\emph{Institut de F\'{i}sica d'Altes Energies (IFAE), Barcelona, Spain }\\
\small\emph{$^f$Institute of Theoretical Sciences, University of Oregon, Eugene, OR97401, USA}
}
}
\begin{document}
\setcounter{page}{0}
\maketitle
\begin{abstract}
In a variety of  well 
motivated  models, such as two Higgs Doublet Models (2HDMs) and the Minimal 
Supersymmetric Standard Model (MSSM),  there are  neutral Higgs bosons that have significantly enhanced couplings to b-quarks and tau leptons in comparison to those of the SM Higgs. These so called non-standard Higgs bosons could be copiously produced at the LHC in association with b quarks, and subsequently decay into b-quark pairs. However, this production channel suffers from large irreducible QCD backgrounds.
We propose a new search strategy for non-standard neutral Higgs bosons at the 7 TeV LHC in the 3b's final state topology.
We perform a simulation  of the signal and backgrounds, 
using state of the art tools and methods for  different sets of selection cuts, and conclude that neutral
Higgs bosons with  couplings
to b-quarks of about 0.3 or larger,  and  masses up to 400 GeV, could be seen with a 
luminosity of 30 $\rm{fb}^{-1}$. In the case of the MSSM we also discuss the complementarity between
the 3b channel and the inclusive tau pair channel in exploring the supersymmetric parameter space.
\end{abstract}
\thispagestyle{empty}
\newpage
\setcounter{page}{1}

\section{Introduction}
\label{sec:Intro}

The origin of electroweak symmetry breaking, leading to the generation of mass 
for the quarks, leptons and the weak gauge bosons
is one of the most outstanding questions in high energy physics.
In the Standard Model (SM), the spontaneous breakdown of the 
electroweak symmetry is induced by the introduction of a scalar doublet field
that acquires a non-vanishing vacuum expectation value.  A physical particle
appears in association with the Higgs mechanism, namely the Higgs boson.  
The tree-level couplings
of quarks and leptons to the Higgs boson are then well defined and are 
proportional to the quark and lepton masses and inversely proportional to the vacuum
expectation value of the Higgs field. Hence, apart from the top quark,
all quarks and leptons have small couplings to the SM Higgs boson.
Therefore the production cross section of a SM Higgs in association with all SM 
fermions, apart from the top quark, is too small to be detectable
at hadron colliders.

It is very likely, however, that the electroweak symmetry breaking sector is 
more complicated than just a single Higgs doublet.  
One of the simplest extensions of  the SM is
the two Higgs doublet model (2HDM).  In this case, quarks  and 
leptons  receive contributions to their masses coming from 
 both Higgs doublets. If the vacuum expectation value of one of the Higgs 
doublets is small, its coupling to some of the quarks can be
very large.  Flavor physics puts however additional constraints on these extended Higgs
sectors: in order to suppress large flavor-changing neutral-current (FCNC) interactions,
 either the coupling of one of the two  Higgs doublets to fermions with a given
electric charge is suppressed, or there is an alignment between  the couplings of the fermions to the two Higgs doublets. The 
so-called Type II 2HDM belongs to the first
class of models: up-type quarks and neutrinos couple 
to one Higgs doublet, $H_u$, and down-type quarks and charged leptons couple to 
the other, $H_d$.

The minimal supersymmetric extension of the SM (MSSM) contains two 
Higgs doublets. At the tree-level, the MSSM Higgs sector is a Type II 2HDM. 
However, fermion couplings to both Higgs doublets are induced at the loop level.
Once Supersymmetry (SUSY) is broken, dangerous FCNC interactions are generated, but since
they are proportional to loop-induced couplings, they tend to be suppressed.
Due to the large top quark mass, the vacuum expectation value
of the Higgs field that only couples to the up sector at the tree level ($H_u$)
cannot be much smaller than the SM one.
Defining $\tan\beta$ as the ratio of the vacuum expectation values of the 
two Higgs fields ($\tan\beta=\langle H_u\rangle/\langle H_d\rangle$), 
this implies that $\tan\beta$ should be of  the order of or larger than 
one. 

In Type II 2HDMs as well as in supersymmetric models, large values of $\tan\beta$ imply a large coupling of the $b$ quark to the non-standard Higgs
bosons, resulting in both a large production cross section of Higgs bosons in association with $b$ quarks as well as a large 
branching fraction of the Higgs bosons decaying into $b$ quarks.
In the supersymmetric case, the precise value of the coupling depends not only on $\tan\beta$, but also on SUSY-breaking effects. 
These effects can modify both the $b$ and $\tau$ couplings to the non-standard Higgs bosons, and hence the branching ratio of these 
Higgs bosons decaying into $b$ quarks and $\tau$ leptons. 

Non-standard Higgs boson production at the LHC has been mainly studied through inclusive
Higgs boson decays into $\tau$ leptons, since this channel has a reasonable signal-to-background ratio.  Currently, the LHC experiments are setting strong bounds~\cite{Atlastautau,CMStautau} on light non-standard neutral Higgs bosons at moderate
or large values of $\tan\beta$, surpassing the previous bounds~\cite{CDFtautau,D0tautau} set by the 
Tevatron experiments.  In this work, we shall study the associated production of a 
non-standard neutral Higgs boson with $b$ quarks at the LHC, with the Higgs
boson subsequently decaying into $b$ quarks. 
 The process involves the production of at least three $b$ quarks in the final 
state and, for large values of $\tan\beta$, the production cross section may be
sizable.  This search channel suffers from a large irreducible background,
since the QCD $b\bar{b}$+X production cross section is much larger than the 
one associated with Higgs production.  Hence, previous experimental studies of this channel
have been mainly performed at the Tevatron~\cite{Abazov:2010ci,Aaltonen:2011nh}, where the 
backgrounds are easier to control, but no CMS or ATLAS analysis is available at present.
This search channel is challenging because the $b$ quark produced in 
association with the Higgs boson typically has low transverse momentum ($p_T$) and 
triggering on such soft $b$ jets, especially for low Higgs boson masses, is difficult
at high instantaneous luminosity. 
After satisfying the trigger requirements, demanding that two of the $b$ jets reconstruct 
the Higgs boson invariant mass helps to improve the signal significance. However 
systematic uncertainties can still be an issue due to the small signal-to-background ratio.
Previous theoretical studies focused on the prospects for the discovery of a non-standard neutral Higgs boson in the $3b$ and $4b$ channels at the 14 TeV LHC~\cite{3bLHCStudies,Kao:2009jv,Baer:2011af}. In particular supersymmetric models such as anomaly-mediated supersymmetry breaking (AMSB) models and gauge-mediated supersymmetry breaking (GMSB) models have been studied~\cite{Baer:2011af}. The aim of our paper is to analyze the reach at the 7 TeV LHC for Higgs bosons arising in generic 2HDMs.
In particular, we determine  the required effective coupling of the neutral Higgs boson to $b$ quarks to have a possible discovery at the 7 TeV LHC with 30 $\rm{fb}^{-1}$ of data. We also consider the specific case of the MSSM and investigate the complementarity to the $A\rightarrow\tau\bar\tau$ searches.

In section 2, we shall present the necessary theoretical background and emphasize the differences between the several two Higgs doublet
extensions discussed above. In section 3 we shall study the reach for non-standard neutral Higgs bosons at the 7 TeV LHC in the $3b$
channel.  We describe our simulation of signal and background and the proposed selection cuts, and then discuss the expected reach in specific 2HDMs, as well as in the MSSM. We reserve section 4 for our conclusions and outlook.

\section{The $\bm{3b}$ channel in 2HDMs}
\label{sec:BMSSM}

In a 
2HDM  the most generic Yukawa couplings of the 
two Higgs doublets with SM quarks and leptons can be written 
as
\be
\mathcal{L}_{\rm{Yuk}} = y_u ~H_u \bar Q U + y_d ~H_d \bar Q D +\tilde y_u ~H_d^\dagger \bar Q U + \tilde y_d ~H_u^\dagger \bar Q D
+ y_\ell ~H_d \bar L E +\tilde y_\ell ~H_u^\dagger \bar L E + h.c.\,,
\ee
in which $H_u$ and $H_d$ are the two Higgs doublets with hypercharge 1/2 and -1/2, respectively.

A generic structure of the four Yukawa couplings leads to Higgs-mediated FCNC interactions already at the tree level. However New Physics (NP) effects in flavor transitions can be reduced by imposing the alignment of up, down and lepton Yukawa couplings~\cite{Pich:2009sp} or, more generically, the Minimal Flavor Violation principle~\cite{D'Ambrosio:2002ex}. 

We shall introduce  the variables $\epsilon_f$, parameterizing the relation between the $y_f$ and $\tilde{y_f}$ couplings in alignment models, 
\be
\tilde y_t = \epsilon_t y_t,~~~ \tilde y_b = \epsilon_b y_d,~~~ \tilde y_\tau = \epsilon_\tau y_\tau\,,
\ee
with generic flavor independent $\epsilon_{t,b,\tau}$ coefficients.
In the following we will focus on the couplings of the three neutral Higgs bosons with the third-generation down-type quarks and leptons.
Assuming that there is no CP violation in the Higgs sector, the couplings of the neutral Higgs bosons with $b$ quarks are given by~\cite{CarenaMrennaW}
\begin{align}
\mathcal{L}_{b} &= \frac{g}{2M_W}\bar{m}_{b} \frac{\tan \beta}{1+\epsilon_{b}\tan\beta}
\left[  A \,i\bar{b}_L b_R \left(1-\frac{\epsilon_b}{\tan\beta}\right)
 + \left(-\frac{\sin\alpha}{\sin\beta} +\epsilon_b \frac{\cos\alpha}{\sin\beta} 
\right) h \bar{b}_L b_R  \right.\nonumber\\
&\ \left. + \left(\frac{\cos\alpha}{\sin\beta} + \epsilon_b 
\frac{\sin\alpha}{\sin\beta} \right) H \bar{b}_L {b}_R + {\rm h.c.} \right]\,,
\end{align}
where $\bar{m}_{b}$ is the running $b$ quark mass and $\alpha$ is the mixing angle between the two scalars $h, H$. The corresponding couplings of the Higgs bosons with the third-generation charged leptons are obtained with the simple exchange $b \leftrightarrow \tau$. 
Hence, in generic aligned 2HDMs the couplings of the pseudoscalar Higgs boson with $b$ quarks and $\tau$ leptons can be parametrized by two independent effective couplings $\tan\beta^b_{\rm{eff}}$ and $\tan\beta^\tau_{\rm{eff}}$

\begin{eqnarray}\label{eq:tanbetab}
\frac{g}{2M_W} \bar{m}_{b}\frac{\tan\beta}{1+\epsilon_{b}\tan\beta}\left(1-\frac{\epsilon_b}{\tan\beta}\right)( A \,i\bar{b} \gamma_5 b)&\equiv& \frac{g}{2M_W} \bar{m}_{b}\tan\beta^b_{\rm{eff}} (A \,i\bar{b} \gamma_5 b)\,,\\\label{eq:tanbetatau}
\frac{g}{2M_W} \bar{m}_{\tau}\frac{\tan\beta}{1+\epsilon_{\tau}\tan\beta}\left(1-\frac{\epsilon_\tau}{\tan\beta}\right)( A \,i\bar{\tau} \gamma_5 \tau)&\equiv& \frac{g}{2M_W} \bar{m}_{\tau}\tan\beta^\tau_{\rm{eff}} (A \,i\bar{\tau} \gamma_5 \tau)\,.
\end{eqnarray}
Moreover, in the decoupling limit, arising at large values of $m_A$ and $\tan\beta$, $\cos\alpha\sim\sin\beta$, $\sin\alpha\sim-\cos\beta$ and consequently the coupling of the heavy scalar $H$ with $b$ quarks ($\tau$ leptons) is also governed by $\tan\beta^b_{\rm{eff}}$ ($\tan\beta^\tau_{\rm{eff}}$). The coupling of the light scalar $h$ is instead SM-like in this limit.

The total production rate of $b$ quarks and $\tau$ pairs mediated by the production of a CP-odd Higgs boson (as well as by the 
heaviest CP-even 
Higgs scalar) in the large $\tan\beta$ regime can be approximated by~\cite{CHWW}

\begin{eqnarray}\label{eq:normalization2HDM}
\sigma(b\bar{b} \rightarrow A) \mathcal{BR}(A\rightarrow b\bar b)
&\sim& \sigma(b \bar{b}h)_{\rm{SM}} (\tan\beta^b_{\rm{eff}})^2\frac{(\tan\beta^b_{\rm{eff}})^2\bar m_b^2 N_c}{(\tan\beta^\tau_{\rm{eff}})^2\bar m_\tau^2+(\tan\beta^b_{\rm{eff}})^2\bar m_b^2 N_c}\,,\\\label{eq:normalization2HDMtau}
\sigma(gg,b\bar{b} \rightarrow A) \mathcal{BR}(A\rightarrow \tau \tau)
&\sim& \sigma(gg,b\bar{b} \rightarrow h)_{\rm{SM}} (\tan\beta^b_{\rm{eff}})^2\frac{(\tan\beta^\tau_{\rm{eff}})^2\bar m_\tau^2}{(\tan\beta^\tau_{\rm{eff}})^2\bar m_\tau^2+(\tan\beta^b_{\rm{eff}})^2\bar m_b^2 N_c}\,,
\end{eqnarray}
where $N_c$ is the number of colors ($N_c=3$) and  $ \sigma(b \bar{b}h)_{\rm{SM}}$ and $\sigma(gg,b\bar{b} \rightarrow h)_{\rm{SM}}$ denote the values of the corresponding SM Higgs boson production cross sections for a Higgs boson of equal mass\footnote{In Eq. (\ref{eq:normalization2HDMtau}) we have neglected the contribution to the production cross section coming from the top-quark loop diagram. The corrections arising from the interference terms between the top-quark and $b$-quark loop diagrams amount only to a few percent~\cite{CHWW}.}.

\bigskip
The MSSM at the tree level is a particular 2HDM of Type II, however at the one-loop level also the Yukawa couplings $\tilde y_u, \tilde y_d,\tilde y_\ell$ are generated. In particular the dominant threshold corrections to the $b$ quark mass are
arising from gluino-sbottom one-loop diagrams and from chargino-stop loops, resulting in 
 $\epsilon_b\sim\epsilon_0 + y_t^2 \epsilon_Y$ with~\cite{Hall:1993gn}--\cite{Carena:1999bh}
\bea\label{eq:epsilon0}
\epsilon_0 &\approx& \frac{2 \alpha_s}{3\pi} M_{\tilde g}\, \mu \, I(M_{\tilde b_1},M_{\tilde b_2},M_{\tilde g}),\\ \label{eq:epsilonY}
\epsilon_Y &\approx& \frac{1}{16\pi^2} A_t\, \mu\, I(M_{\tilde t_1},M_{\tilde t_2},\mu),
\eea
in which $M_{\tilde g}$ is the gluino mass, $M_{\tilde b_i}$ and $M_{\tilde t_i}$ the sbottom and stop masses and $\mu$ the 
Higgsino mass parameter. $A_t$ and $y_t$ are the top trilinear term and the top
Yukawa coupling, respectively.

Similarly, the corrections to the $\tau$ mass are dominated by wino and bino exchange contributions that are usually small, since they are suppressed by 
the electroweak coupling and have the form~\cite{Dobrescu:2010mk,Altmannshofer:2010zt}

\be
\epsilon_\tau\approx\frac{3\alpha_2}{8\pi}\mu\, M_2\,I(M_{\tilde \nu_\tau},M_{2},\mu),
\ee
with $M_2$ the wino mass, $M_{\tilde \nu_\tau}$ the sneutrino mass.

The effective couplings of the 
CP-odd and heavier CP-even Higgs bosons 
of the MSSM with $b$ quarks and $\tau$ leptons are then given by 
Eqs.~(\ref{eq:tanbetab})~{and}~(\ref{eq:tanbetatau}) with the resummation factors $\epsilon_b$ and $\epsilon_\tau$ given just above. 
As a result, contrary to generic aligned 2HDMs, in the MSSM the two couplings will be uniquely determined, once the supersymmetric spectrum is specified.

\section{Early LHC prospects for the $\bm{3b}$ channel}
\label{sec:LHC}
\subsection{Simulation of Signal and Background}

Signal and background processes are modeled using the MadEvent5~\cite{Alwall:2011uj} event 
generator interfaced with Pythia 6.4~\cite{Sjostrand:2006za} for parton showering and hadronization,
using a matrix-element parton-shower matching algorithm to avoid double-counting of partonic
configurations. We allowed for up to two additional partons in the final state of the hard process when 
following the  shower-$k_T$  scheme outlined in Refs.~\cite{Alwall:2007fs,Alwall:2008qv}, with $k_T$-matching 
scale of $30$~GeV.  These samples were generated for $pp$ collisions at $\sqrt{s} = 7$~TeV using the 
CTEQ6L1 parton distribution functions (PDFs)~\cite{Pumplin:2002vw}. 

The QCD production of multiple heavy quarks is the main source of background.
We generated two separate QCD background samples: a $b\bar{b}j+X$ ($j=u,d,s,c,g$) sample in 
which the additional partons used in matching to the parton-shower are light or charm quarks and a
"$3b$" ($b\bar{b}b+X$ and $b\bar{b}\bar{b}+X$) sample in which the additional partons could be either light or 
heavy quarks.\footnote{
This separation of QCD background into the $bbj$ and $3b$ samples does not 
model $b$ jets with $p_T$ below $\sim 40$~GeV very well. However once we impose $p_T$ cuts on the 
jets as described in the following, the effects are only at the $\sim 10$\% level.} 

We cluster particle jets using the anti-$k_T$  algorithm implemented in FastJet-2.4.3~\cite{Cacciari:2008gp} with 
a radius parameter $R = 0.4$. To better simulate the experimental $b$ jet energy resolution, we exclude
neutrinos from jet clustering, which in the case of semileptonic $b$ decays can carry away a significant 
fraction of energy. Furthermore, we apply a jet energy smearing of $100\%/\sqrt{E/{\rm GeV}}$ to model the
typical calorimeter energy resolution of LHC experiments.

Since the signal typically contains three $b$ quarks, flavor tagging becomes an effective tool to suppress QCD multijet backgrounds. We assume a constant $b$-tagging efficiency of $60$\%, a $c$-jet mis-tag rate of $10$\% and a light-jet 
mis-tag rate of $1$\%~\cite{Aad:2009wy}.  This choice can be considered conservative, as LHC experiments have already
developed sophisticated $b$-tagging algorithms~\cite{atlasbtagging} exceeding the performance assumed in this
paper.  The low mis-tag rate of $c$- and light-jets leads to the $bbj$ and $3b$ backgrounds being comparable once
three $b$-tagged jets are required.

We consider two sets of event selection criteria:
\begin{enumerate}
\item Selection I: events are required to have exactly three $b$-tagged jets with $p_T > 60$~GeV and $|\eta| < 2.0$. 
\item Selection II: events are required to have exactly three $b$-tagged jets with $p_T > 50$~GeV and $|\eta| < 2.0$,
         and the leading $b$-tagged jet to have $p_T> 130$~GeV.
\end{enumerate}

In both cases jets are required to be relative central to ensure they are contained within the tracker volume
and can therefore be tagged with high efficiency~\cite{Aad:2009wy}.

The high instantaneous luminosities delivered by the LHC has forced to raise threshold in the trigger menus in ATLAS and CMS. As a result, the first selection is somewhat optimistic in that the jet $p_T$ requirements may be too low for these events to
satisfy trigger requirements with high efficiency. For instance, events satisfying
Selection I would have a low efficiency to satisfy the 2$b$/2j ATLAS trigger~\cite{atlasbjettrigger} requirements, 
which appears as one of the most suitable unprescaled triggers for this topology. On the other hand, Selection II would be more representative of the kind of minimum jet $p_T$ requirements
applied by the 2$b$/2j ATLAS trigger once trigger turn-on effects are considered. Nevertheless, we consider
Selection I to explore the potential sensitivity gains at low $m_A$ values, which could motivate designing an
optimized trigger strategy for such lower -$p_T$ events at the LHC experiments.

\begin{figure}[t]
\center
\includegraphics[width=0.325\textwidth]{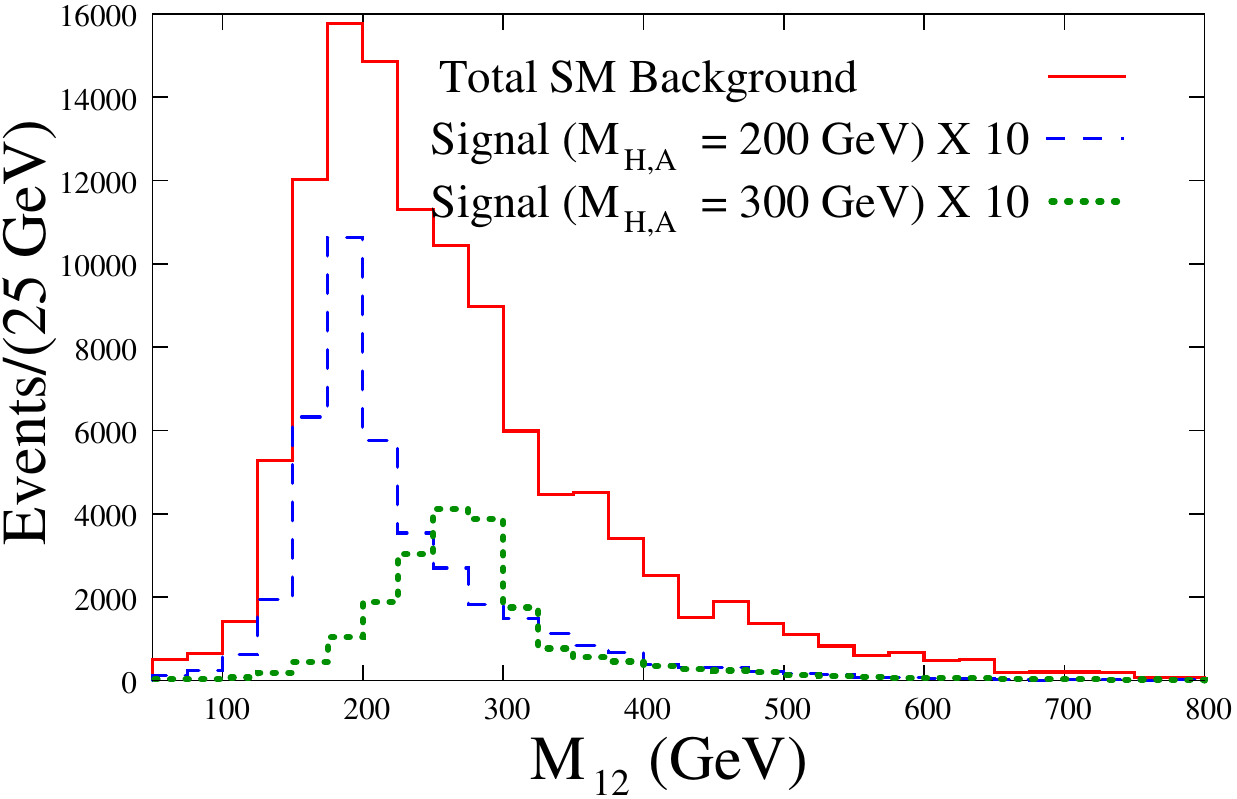}
\includegraphics[width=0.325\textwidth]{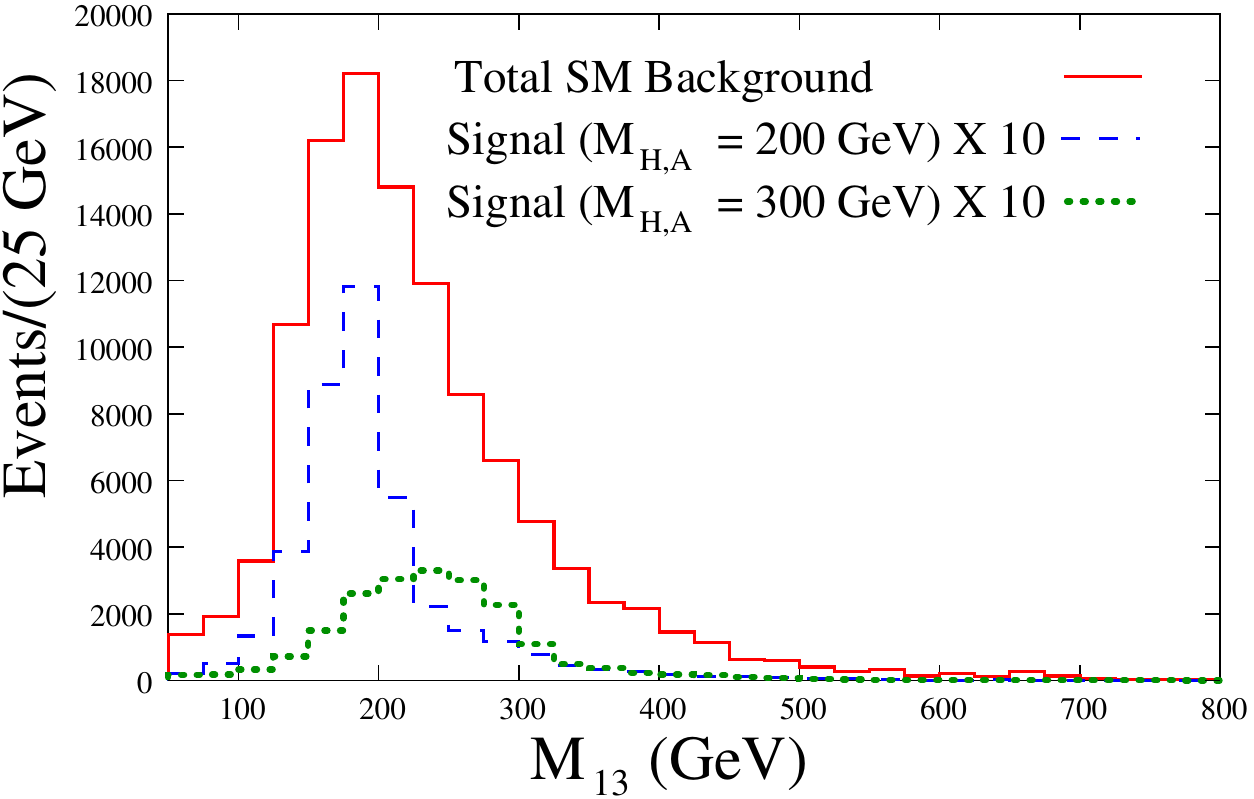}
\includegraphics[width=0.325\textwidth]{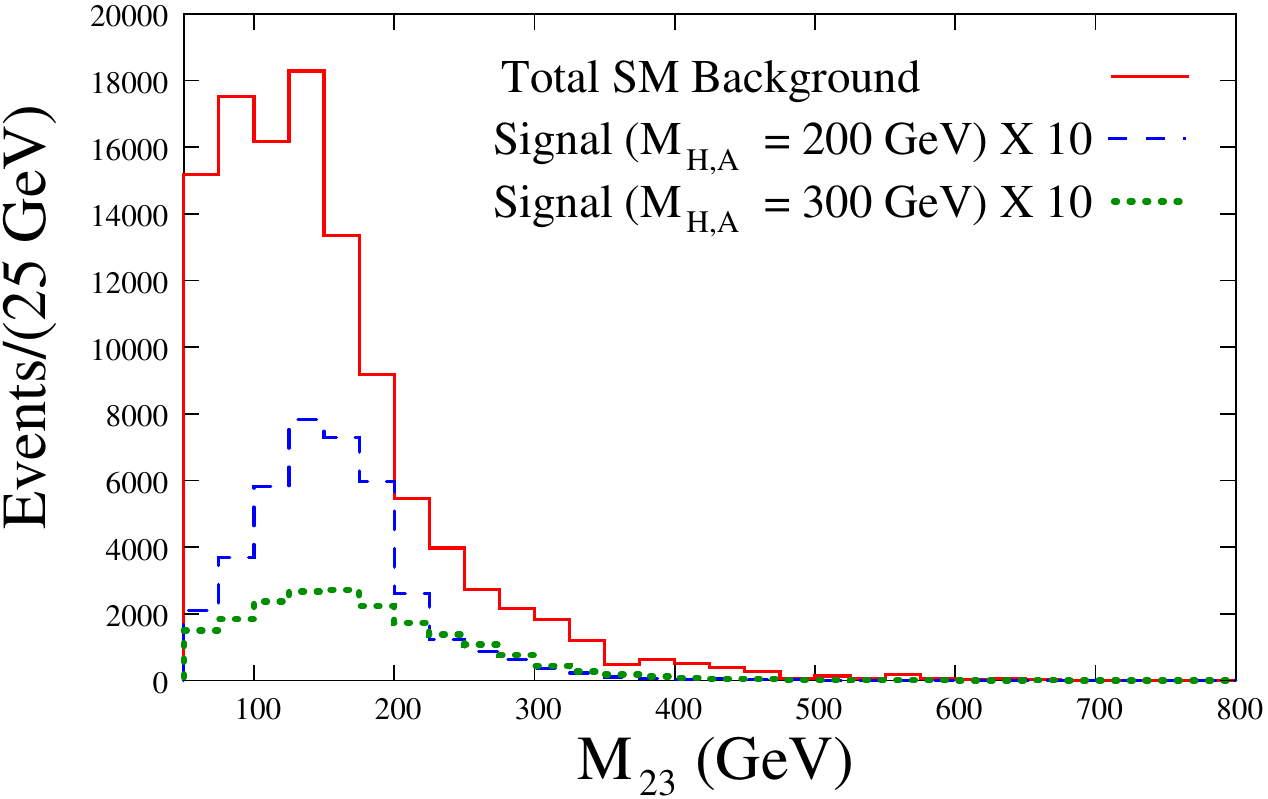}
\caption{ \small Invariant mass distributions for each of the $b$-jet pairs for events satisfying Selection I and assuming an
integrated luminosity of 30 fb$^{-1}$ in $pp$ collisions at $\sqrt{s}=7$~TeV. The expected distribution for the total
background (red histogram) is compared to that for a signal with $m_A=200$~GeV (blue histogram) and $m_A=300$~GeV (green histogram) assuming $\tan\beta=30$. The signal expectation has been scaled by a factor 
of ten for visibility.}
\label{Fig:Distributions}
\end{figure}

In most part of the parameter space under consideration, there is no obvious way to accurately distinguish the pair of $b$ jets coming from the decay of the heavy scalar and the $b$ jet produced in association with it. 
However Selection II with masses $m_A \lsim 260$~GeV represents an exception: the $b$ jet produced in
association with the Higgs boson is the leading one ($b_1$) because the $b$ jets coming from 
the decay of the Higgs are too soft to satisfy the selection criteria. Therefore in Selection II
for low $m_A$ the invariant mass of the second-leading $b$ jet ($b_2$) and the third-leading $b$ jet ($b_3$)  
reconstructs the Higgs mass, while for large $m_A$ ($m_A \gsim 
260$~GeV) the invariant mass
of $b_1$ with $b_2$ or $b_3$ reconstructs the Higgs mass. Nevertheless to improve the acceptance, we consider all three possible 
combinations of $b$-jet pairs and require that  the invariant mass of 
at least one of them is within a window around the peak of the Higgs boson invariant mass distribution.
Figure~\ref{Fig:Distributions} compares the invariant mass distribution between signal and background for
each of the possible $b$-jet pairs. As it can be appreciated, the signal distribution consistently peaks
at values lower than the physical mass of the Higgs boson due to a combination of the jet smearing\footnote{We checked that, in the range of $\tan\beta$ we are considering for our analysis, the effects on the b-pair invariant mass distributions coming from the physical width of the Higgs are negligible, if compared to the width of the invariant mass distribution coming from jet smearing.} and
energy loss via neutrinos, with the PDF suppression involved in producing such a massive resonances. 
This effect becomes more significant for heavier resonance. If an excess in the invariant mass distribution of a pair of b quarks is discovered at the LHC, one needs to extract the actual mass of the resonance through proper simulation.  The chosen central value of the mass window is
shown in Table~\ref{tab:mass_window} for each of the $m_A$ values considered.

Additionally, we studied the effect on varying the width of
the mass window about the peak and found that a typical width of 
$|M_{bb} - m_{\rm peak}|\leq 25$~GeV yields good results across the whole mass range.
Increasing the width of the mass window for heavier masses and reducing
the width for lower masses can lead to a improvement in the significance of only of a
few percent. Finally, we checked that imposing a cut on $\Delta R$ between any two $b$ jets
did not lead to a marked improvement in the signal significance.
\begin{table}[h!]
\centering
\begin{tabular}{c|cccccc}
$m_A$ (GeV) & 150 & 200 & 250 & 300 & 400 & 500 \\
\hline
$m_{\rm peak}$ (GeV) & 150  &190  & 230 & 250  & 350 & 450 
\end{tabular}
\caption{Physical masses, $m_A$, used in our signal samples, and the corresponding central values of the mass window. The mass window used in the signal selection is chosen to be  $|M_{bb} - m_{\rm peak}|\leq 25 $~GeV. \label{tab:mass_window}  }
\centering
\end{table}

\subsection{Prospects and Significance}

As discussed in Sec.~\ref{sec:BMSSM}, in a generic 2HDM the couplings of the $A,H$ bosons with $b$ quarks and with $\tau$ leptons are independent and parametrized by the
effective couplings $\tan\beta_{\rm eff}^b$ and $\tan\beta_{\rm eff}^\tau$, respectively.
As shown by Eq. (\ref{eq:normalization2HDM}), the cross section times branching ratio for
non-standard neutral Higgs bosons produced in association with a $b$ quark which subsequently decay into
a pair of $b$ quarks  
has only a mild dependence on the choice of $\tan\beta_{\rm eff}^\tau$. 
Without lost of generality we fix $\tan\beta_{\rm eff}^\tau=5$ to be in agreement with the 
present bounds coming from LHC $A,H\rightarrow\tau\bar\tau$ searches~\cite{Atlastautau,CMStautau}.

In addition, we are focusing on the parameter region with sizable $\tan\beta_{\rm eff}^b$ and moderate values of  $m_A$.  
In this case,   the heavy CP-even scalar $H$ and the pseudoscalar $A$ can only be slightly split in mass, so that
the two Higgs bosons will appear at the LHC in the same resonance region with combined cross section.
Hence, for our analysis, the only two relevant free parameters are $m_A \sim m_{H}$ and $\tan\beta_{\rm eff}^b$. 
To compute the rate of the signal, we double the cross section obtained for the CP-odd Higgs and use the narrow width approximation which is valid 
in the entire mass range we consider as long as $\tan\beta_{\rm eff}^b$ is not too large ($\tan\beta_{\rm eff}^b\lesssim 80$). 
\begin{table}[h!] 
\addtolength{\arraycolsep}{10pt}
\renewcommand{\arraystretch}{1.3}
\centering
\begin{tabular}{|c|c|c|c|}
\hline\hline
& {\footnotesize $3b$} & {\footnotesize $bbj$} & {\footnotesize Signal ($m_A$ in GeV)}   \\ 
&                    &                            &~150 ~~~~~   200~~~~~~~   250 ~ ~~~300  ~~ 400~~~   500\\
\hline\hline
{\footnotesize After matching $/ 10^3$  }              & $4800$ & $2.2\times 10^6$ &$420$ ~~~~~~~$180$ ~~~~~~~$90$ ~~~~~~~$45$~~~~~~ $14$~~~~~~$5$\\ 
{\footnotesize Selection I} & $45000$ & $69000$ &~$3900$ ~~~~ $4500$ ~~~~ $3600$ ~~~ $2350$~~~~$960$~~~$150$\\ 
\hline\hline
{\footnotesize $m_A=150$ GeV }& $21000$ & $33000$ &~ 3300~~~~~~~~~~~~~~~~~~~~~~~~~~~~~~~~~~~~~~~~~~~~~~~~~~~~ \\ 
{\footnotesize $m_A=200$ GeV} & $24000$ & $39000$&       ~~~~~~~~~~~ 3600~~~~~~~~~~~~~~~~~~~~~~~~~~~~~~~~~~~~~~~  \\ 
{\footnotesize $m_A=250$ GeV} & $19000$ & $30000$ &    ~~~~~~~~~~~~~~~~~~ ~~      2500~~~~~~~~~~~~~~~~~~~~~~~~~~ \\ 
{\footnotesize $m_A=300$ GeV} & $16000$ & $26000$ &       ~~~~~~~~~~~~~~~~~~~~~~~~~~~~~      1500~~~~~~~~~~~\\ 
{\footnotesize $m_A=400$ GeV} & $6300$ & $9300$ &      ~~~~~~~~~~~~~~~~~~~~~~~~~~~~~~~~~~ ~~~     420~\\ 
{\footnotesize $m_A=500$ GeV} & $2400$ & $3300$ &       ~~~~~~~~~~~~~~~~~~~~~~~~~~~~~~~~~~~~~~~~~~~~~~~~~~~~~~      60\\ 
\hline\hline
\end{tabular}
\caption{\small Expected number of background and signal (at $\tan \beta_{\rm eff}^b = 30$) events per 30 $\rm{fb}^{-1}$ of data at the 7 TeV LHC, after imposing Selection I presented in the text (above double line) and after 
the mass window selection presented in Table~\ref{tab:mass_window} (below double line). 
The first row shows the total events in each channel before event selection criteria are imposed.
}
\label{tab:crossSections}
\end{table}
\begin{table}[h!] 
\addtolength{\arraycolsep}{10pt}
\renewcommand{\arraystretch}{1.3}
\centering
\begin{tabular}{|c|c|c|c|}
\hline\hline
& {\footnotesize $3b$} & {\footnotesize $bbj$} & {\footnotesize Signal ($m_A$ in GeV)}   \\ 
&                    &                            &~150 ~~~~~   200~~~~~~~   250 ~ ~~~300  ~~ 400~~~   500\\
\hline\hline
{\footnotesize After matching $/ 10^3$  }              & $4800$ & $2.2\times 10^6$ &$420$ ~~~~~~~$180$ ~~~~~~~$90$ ~~~~~~~$45$~~~~~~ $14$~~~~~~$5$\\ 
{\footnotesize Selection II} & $24000$ & $42000$ &~~$1200$ ~~~ $1650$ ~~~~ $2100$ ~~~~ $1850$~~~~~$850$~~~$120$\\ 
\hline\hline
{\footnotesize $m_A=150$ GeV }& $6300$ & $11000$ & 800~~~~~~~~~~~~~~~~~~~~~~~~~~~~~~~~~~~~~~~~~~~~~~~~~~~~ \\ 
{\footnotesize $m_A=200$ GeV} & $10000$ & $19000$&       ~~~~~~~~~~~ 1350~~~~~~~~~~~~~~~~~~~~~~~~~~~~~~~~~~~~~~~  \\ 
{\footnotesize $m_A=250$ GeV} & $12000$ & $20500$ &    ~~~~~~~~~~~~~~ ~~~~~~      1700~~~~~~~~~~~~~~~~~~~~~~~~~ \\ 
{\footnotesize $m_A=300$ GeV} & $11000$ & $20000$ &       ~~~~~~~~~~~~~~~~~~~~~~~~~~~~~~      1200~~~~~~~~~~~\\ 
{\footnotesize $m_A=400$ GeV} & $4800$ & $9000$ &      ~~~~~~~~~~~~~~~~~~~~~~~~~~~~~~~~~~~~ ~~~     390\\ 
{\footnotesize $m_A=500$ GeV} & $1900$ & $2900$ &       ~~~~~~~~~~~~~~~~~~~~~~~~~~~~~~~~~~~~~~~~~~~~~~~~~~~~~~      45\\ 
\hline\hline
\end{tabular}
\caption{\small Expected number of background and signal (at $\tan\beta_{\rm eff}^b = 30$) events per 30 $\rm{fb}^{-1}$ of data at the 7 TeV LHC, after imposing Selection II presented in the text (above double line) and after 
the mass window selection presented in Table~\ref{tab:mass_window} (below double line). 
The first row shows the total events in each channel before event selection criteria are imposed.
}
\label{tab:crossSections2}
\end{table}


\begin{table}[h!] 
\addtolength{\arraycolsep}{10pt}
\renewcommand{\arraystretch}{1.3}
\centering
\begin{tabular}{|c|c|c||c|c|}
\hline\hline
&\multicolumn{2}{|c|}{Selection I }& \multicolumn{2}{|c|}{Selection II }\\
\hline
  & {\footnotesize $S/B$ }&{\footnotesize $S/\sqrt B$ }&{\footnotesize $S/B$ }&{\footnotesize $S/\sqrt B$ }\\
  \hline
  {\footnotesize $m_A=150$ GeV }& 0.06 & 14.1 & 0.047 & 6.2\\
  {\footnotesize $m_A=200$ GeV }& 0.057 & 14.4 & 0.048 & 7.9\\
  {\footnotesize $m_A=250$ GeV }& 0.051 & 11.4 & 0.052 & 9.4\\
  {\footnotesize $m_A=300$ GeV }& 0.035 & 7.3 & 0.038 & 6.8\\
  {\footnotesize $m_A=400$ GeV}& 0.027 & 3.4 & 0.028 & 3.3\\
  {\footnotesize $m_A=500$ GeV }& 0.01 & 0.8 & 0.01 & 0.7\\
  \hline\hline
\end{tabular}
\caption{\small Signal (at $\tan \beta_{\rm eff}^b = 30$) to background ratio and significance $S/\sqrt B$ per 30 $\rm{fb}^{-1}$ of data at the 7 TeV LHC, using the two
Selections presented in the text.
}
\label{tab:significance}
\end{table}

In Tables~\ref{tab:crossSections} and~\ref{tab:crossSections2} we present
the number of signal and background events per 30 $\rm{fb}^{-1}$ in each of the
test mass windows, assuming $\tan\beta_{\rm eff}^b=30$. In Table~\ref{tab:significance}, we compare the signal statistical local significance\footnote{Note that in Table~\ref{tab:significance} and in the rest of the paper we are only presenting the local significance for a Higgs with mass in one of the selected mass windows. The study of the lookelsewhere effect goes beyond the scope of this paper.} from applying Selection I with that of Selection II, assuming an integrated luminosity of 30 fb$^{-1}$. We can see that, as expected, Selection I has a markedly better statistical sensitivity for $m_A<300$~GeV.

\begin{figure}[h!]
\center
\includegraphics[width=0.45\textwidth]{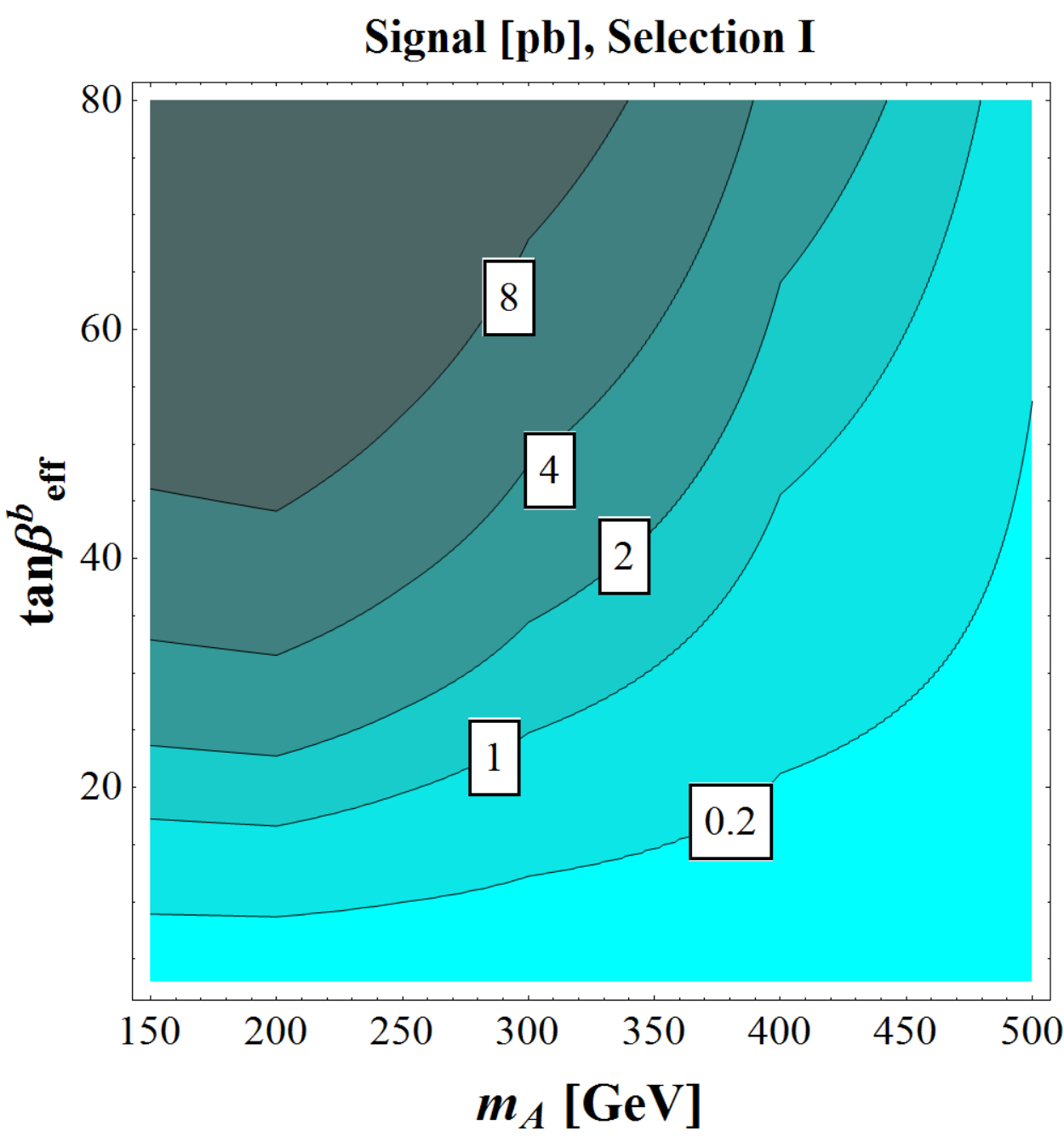}~~~
\includegraphics[width=0.45\textwidth]{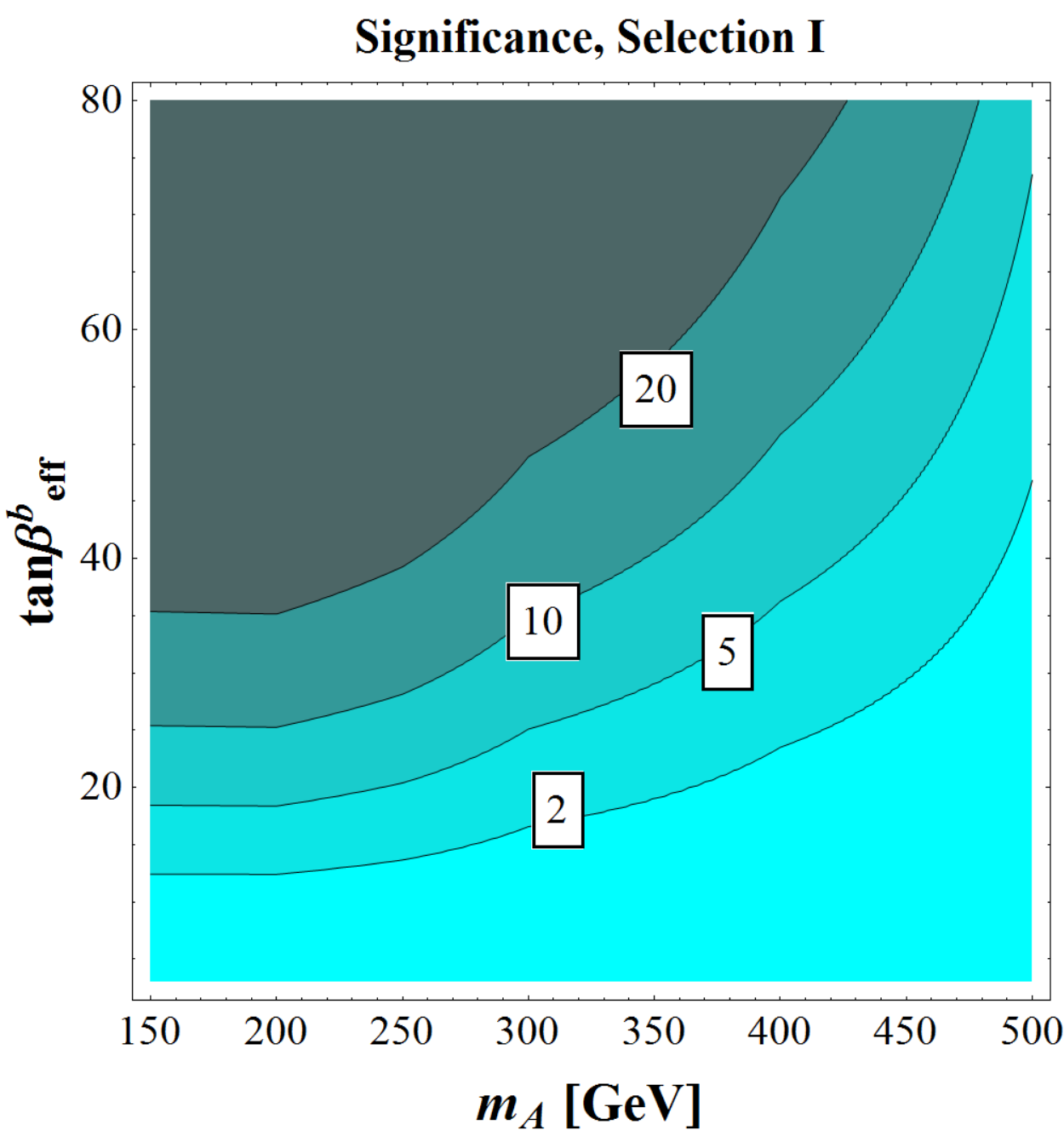}\vspace{0.5cm}
\includegraphics[width=0.45\textwidth]{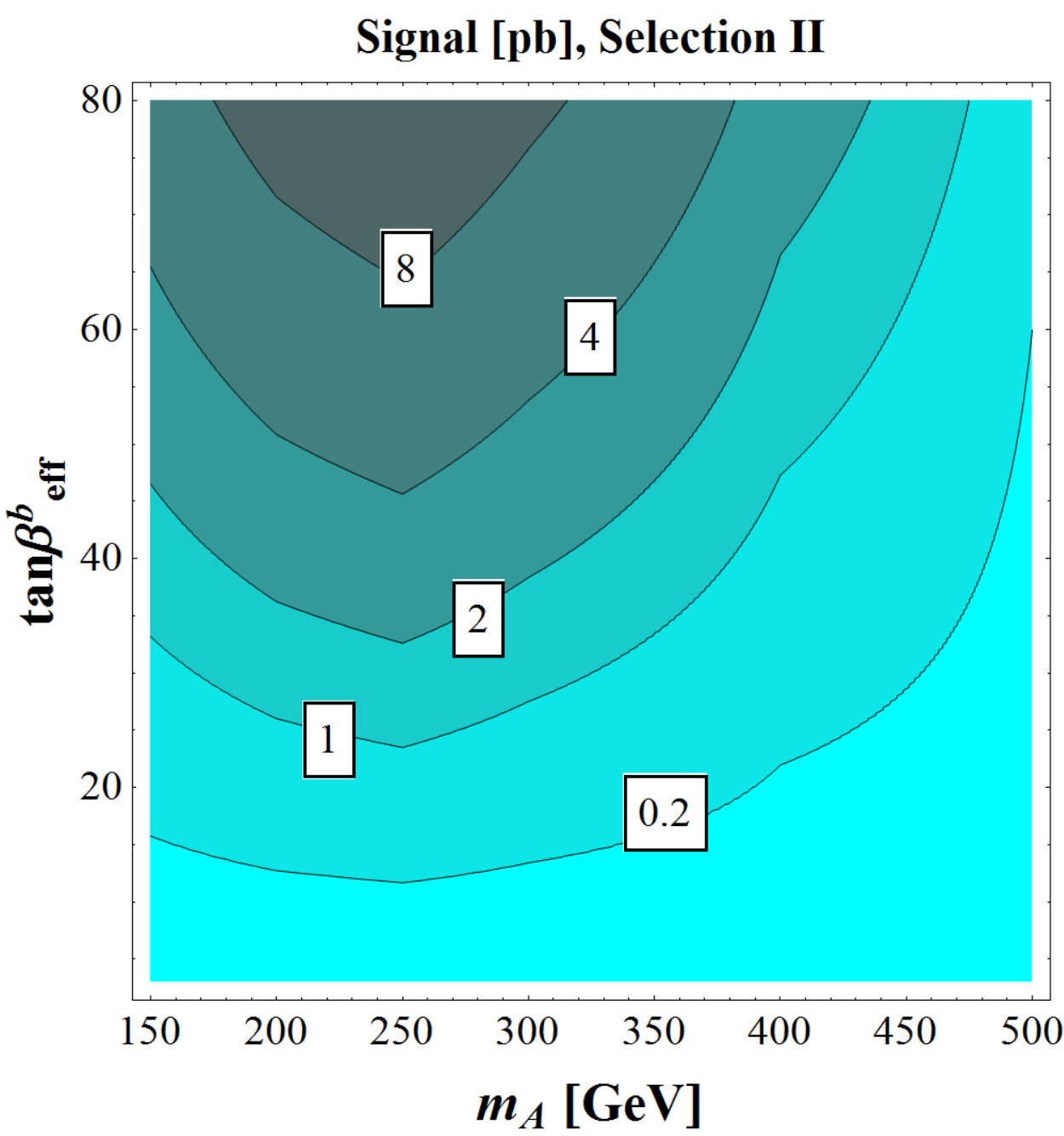}~~~
\includegraphics[width=0.45\textwidth]{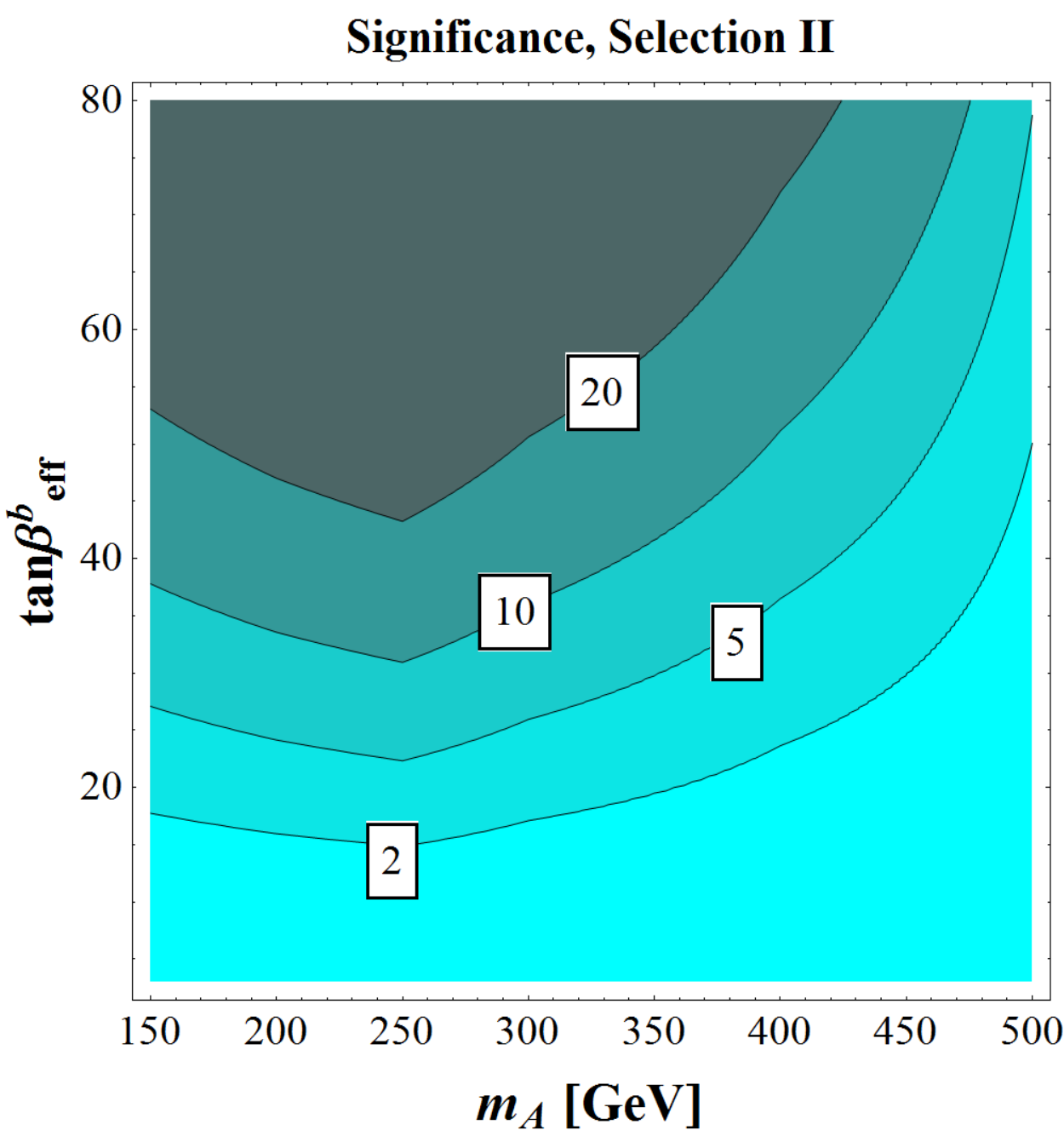}
\caption{ \small Accepted signal cross section in pb (left column) and statistical significance $S/\sqrt{B}$ (right column) in a generic 2HDM (assuming $\tan\beta_{\rm eff}^\tau=5$ ) at the 7 TeV LHC for an integrated luminosity of 30 $\rm{fb}^{-1}$.  In the top row, we show the results after imposing Selection I and the mass window cuts detailed  in Table~\ref{tab:mass_window}. In the bottom row, the results from applying Selection II and the mass window cuts presented in Table~\ref{tab:mass_window}. }
\label{Fig:Summary2HDM}
\end{figure}

Using Eq. (\ref{eq:normalization2HDM}), it is straightforward to
generalize these results to different values of $\tan\beta_{\rm eff}^b$. 
In Fig.~\ref{Fig:Summary2HDM} we present the accepted signal cross section, after that all cuts are implemented, and the statistical significance at 
30~$\rm{fb}^{-1}$ LHC as a function of the two free parameters of the theory, 
$\tan\beta_{\rm eff}^b$ and $m_A$ for both Selection I and Selection II. The results are encouraging. 
For a 7 TeV LHC run with a total integrated luminosity of 30 fb$^{-1}$, we can probe a large parameter region of the 2HDM by searching for heavy Higgs scalars in the $3b$ final state.  For example, applying 
Selection I,  a coupling of the pseudoscalar with $b$ quarks of the order $\sim 0.3$, and hence $\tan\beta_{\rm eff}^b\sim 30$, could lead to a 5$\sigma$ significance for a Higgs boson mass up to $\sim 370$ GeV, with 30 $\rm{fb}^{-1}$ of data.  However, from Table 4, we also see that for moderate $\tan \beta_{\rm eff}^b$, the final $S/B\lesssim 0.1$. Therefore the potential systematic uncertainties which have not been accounted for in our
analysis could make signal identifications challenging. We expect that a detailed experimental analysis exploiting the sideband regions in the invariant mass distribution to constrain systematic uncertainties in the background prediction can nevertheless achieve a high sensitivity. In any case, for larger values of $\tan \beta_{\rm eff}^b (\sim 60)$, and hence a coupling $\sim 0.6$, we can achieve a  significance $\gsim 10$ with 30~fb$^{-1}$ of data in almost the entire range of masses considered. For such large effective 
couplings also the signal-to-background ratio would be more favorable, $S/B\lesssim 1/5$.

Comparing the first and second rows of Fig.~\ref{Fig:Summary2HDM}, we note that there is a difference in shape of the contours. Applying Selection II, the best reach is for $m_A \sim 250$ GeV. On the other hand, the reach when using Selection I monotonically decreases as the value of $m_A$ increases, which is the result of the rapid decrease in signal rate.   
This difference in the shapes of the accepted signal cross section and of the statistical significances are related to the
different cuts on the three highest $p_T$ jets.  First of all, the $b$ jet produced in association with the Higgs boson has a steeply falling distribution, suppressed by the PDFs.  Therefore, it is unlikely that this jet can satisfy the cut of $p_T > 130$ GeV on the leading jet in Selection II. This explains why Selection I leads to better sensitivity, in particular in the low mass region.   On the other hand, for $m_A \sim 250$ GeV, it is easier for the $b$ jets from Higgs decay to be the leading jet and to satisfy this cut. 
This explains why the reach in Selection II is better for higher Higgs masses than for lower ones.
This effect suggests that an asymmetric jet energy cut, similar to that of Selection II, could have advantages. 
The $p_T$ cut on the leading jet could be optimized further, such as requiring it to be proportional to the target 
signal mass.

\bigskip
The possibility of detecting the pseudoscalar and the heavy scalar of the MSSM in the $3b$ channel deserves a special discussion, since in the MSSM the effective couplings $\tan\beta^b_{\rm{eff}}$ and $\tan\beta^\tau_{\rm{eff}}$ defined in Eqs.~(\ref{eq:tanbetab})~and~(\ref{eq:tanbetatau}) are determined, once the SUSY spectrum is specified.

As discussed in Sec.~\ref{sec:BMSSM}, contrary to generic 2HDMs, in the MSSM the coupling of the pseudoscalar Higgs with
$b$ quarks and $\tau$ leptons depend equally on $\tan\beta$ but have a different dependence on corrections arising at the one-loop level.
Typically, for gluinos at the TeV scale, stops, sbottoms and charginos at a few hundred GeV and $A_t$ of the order 1-2 TeV, $\epsilon_b$ is at the few $\%$ level. 
On the other hand, in the lepton sector typically $\epsilon_\tau \sim \mathcal{O}(10^{-3}).$\footnote{Notice, however, that scenarios with light third-generation sleptons and large values of the $\mu$ parameter can {also} lead to values of $\epsilon_\tau$ at the few $\%$ level~\cite{Carena:2011aa}.}

\begin{figure}[h!]
\center
\includegraphics[width=0.45\textwidth]{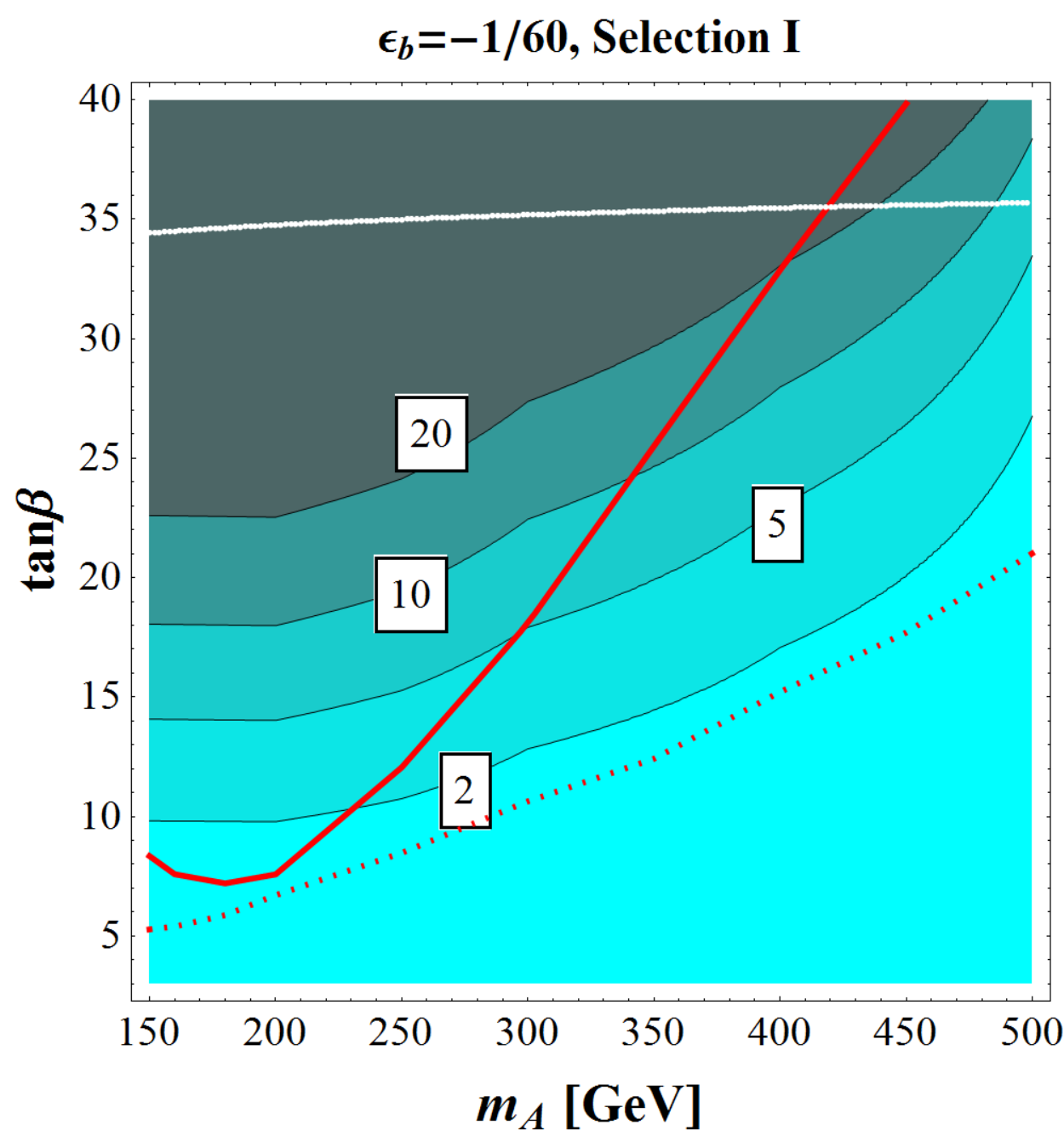}~~~
\includegraphics[width=0.45\textwidth]{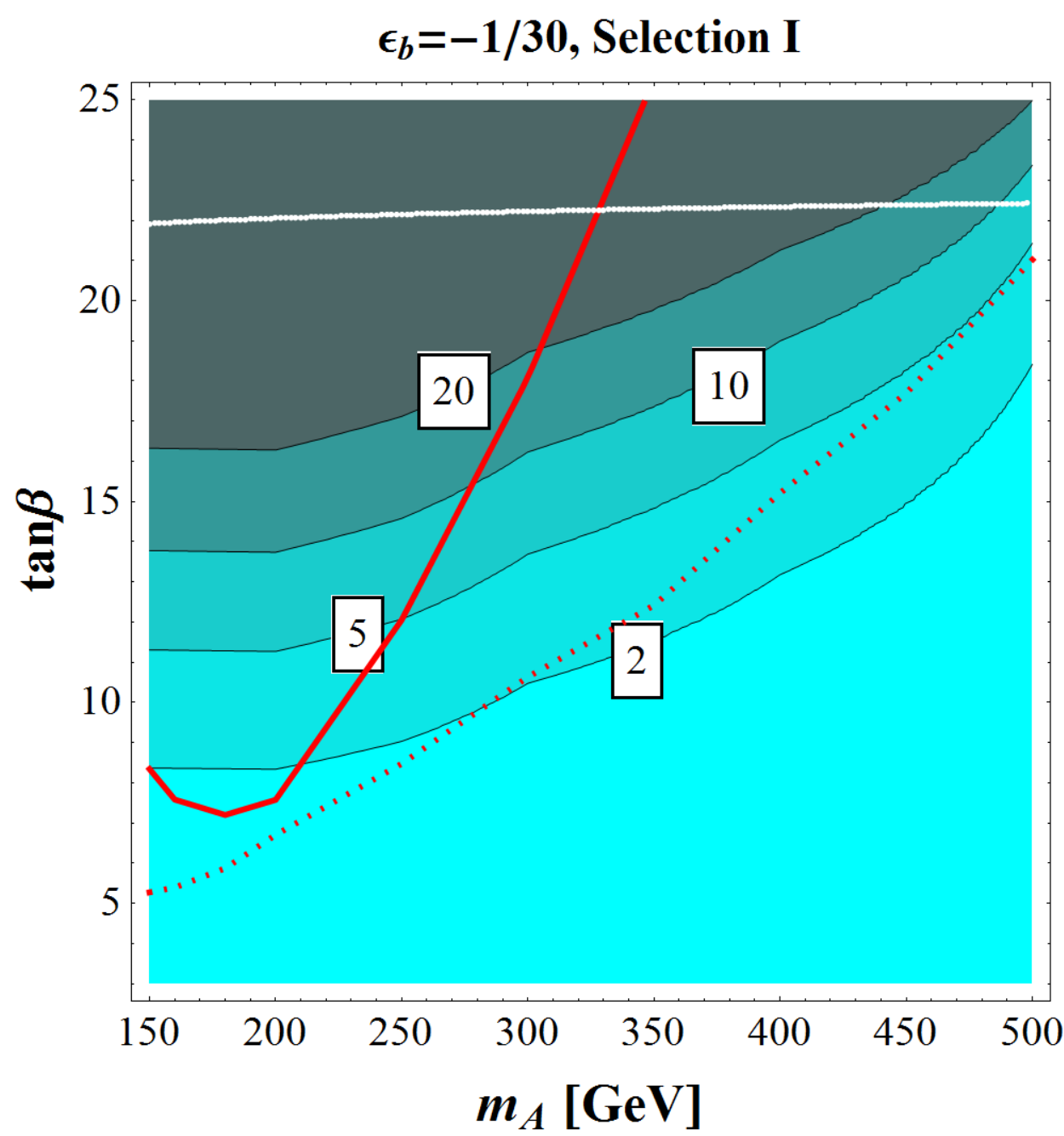}\vspace{0.5cm}
\includegraphics[width=0.45\textwidth]{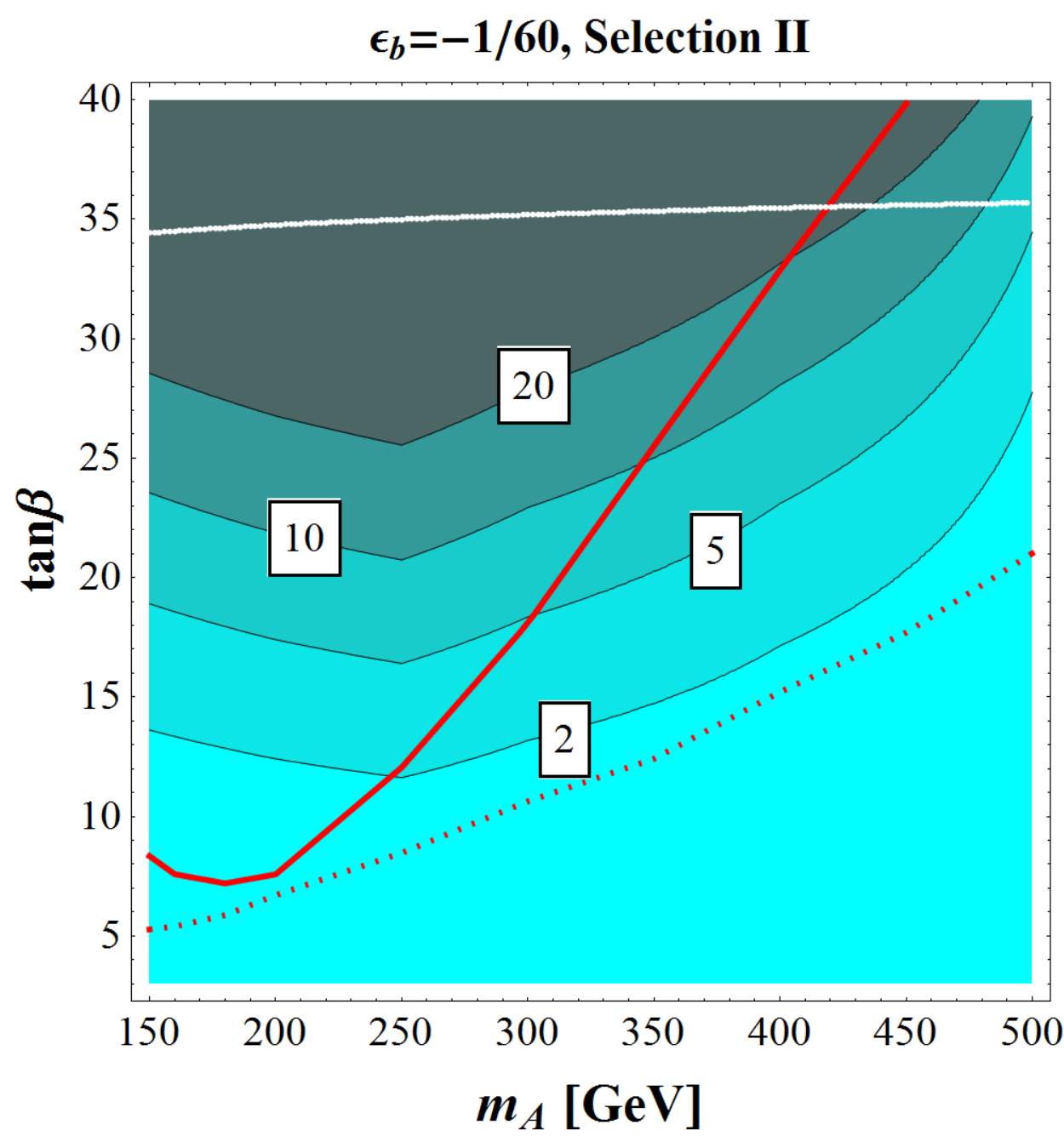}~~~
\includegraphics[width=0.45\textwidth]{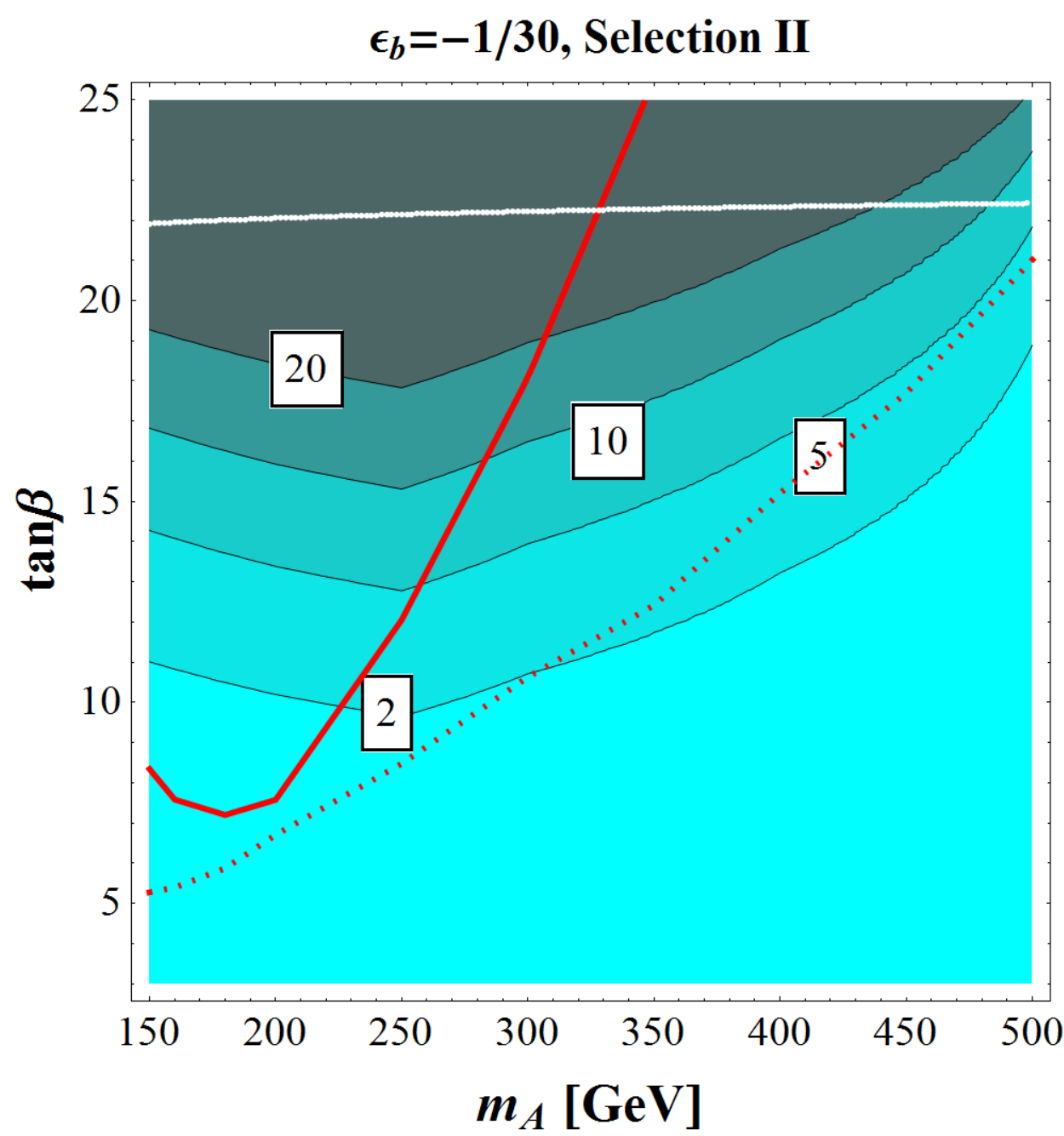}
\caption{ \small Statistical significance at the 7 TeV LHC for an integrated luminosity of 30 $\rm{fb}^{-1}$ in two different scenarios:  $\epsilon_b=-1/60$ (left panels) and $\epsilon_b=-1/30$ (right panels). The results for Selection I and II are
shown in the top and bottom rows, respectively. The red solid (dashed) line represents the present (projected at 30 fb$^{-1}$) bound on non-standard Higgs bosons decaying to $\tau\bar\tau$~\cite{CMStautau}. The area below the white solid line corresponds to the region of validity of the narrow-width approximation ($\Gamma_A\lesssim m_A/10$).}
\label{Fig:SummaryI}
\end{figure}

In our numerical analysis, we choose two 
representative scenarios: the first with $\epsilon_\tau =0$ and $\epsilon_b =-1/60$, and the second with $\epsilon_\tau =0$ and $\epsilon_b =-1/30$. Both scenarios can be achieved in models with a large and negative $\mu$ term (see Eqs. (\ref{eq:epsilon0}) and (\ref{eq:epsilonY})). 
The effects of introducing a small but non-zero $\epsilon_\tau$ will not  significantly modify our conclusions. 
These scenarios
are presented in Fig.~\ref{Fig:SummaryI}. The plots on the left represent the case $\epsilon_\tau=0$ and 
$\epsilon_b=-1/60$; the ones on the right 
 $\epsilon_\tau=0$ and $\epsilon_b=-1/30$. In white we present the 
bound on $\tan\beta$ coming from the requirement that the narrow width 
approximation is valid ($\Gamma_A\lesssim \frac{m_A}{10}$).

It is interesting to compare the shape of the exclusion bound
from the LHC $A\rightarrow\tau\bar\tau$ 
search to the one of the constant 
significance contours for the $A\rightarrow b\bar b$ channel. 
For large values of $m_A$, the present CMS bound~\cite{CMStautau} obtained with $\sim 5\, \rm{fb}^{-1}$ of data is weaker than what was expected. As a consequence, the slope of the present $\tau\bar\tau$ exclusion bound (solid red line in the figure) is much steeper than the slope of the $A\rightarrow b\bar b$ constant significance contours, at large values of $m_A$. Differently, the slope of the expected bound projected at 30 fb$^{-1}$ (dashed red line in the figure) gets much closer to the slope of the $A\rightarrow b\bar b$ constant significance contours, especially at small values of $\epsilon_b$ ($\epsilon_b=-1/60$). For larger $\epsilon_b$ ($\epsilon_b=-1/30$) one can still note a difference in the slopes, that is mainly due to the different scaling in $\tan\beta$ of $pp\rightarrow b\bar b A$ with $A\rightarrow b\bar b$ and $pp\rightarrow b\bar b A$ with $A\rightarrow \tau\bar\tau$, as shown by Eqs. (\ref{eq:normalization2HDM}), (\ref{eq:normalization2HDMtau}) once $\tan\beta^b_{\rm{eff}}$ and $\tan\beta^\tau_{\rm{eff}}$ are replaced with their expressions in (\ref{eq:tanbetab}), (\ref{eq:tanbetatau}) and $\epsilon_\tau$ is set to 0. Thanks to this different scaling, the $3b$ channel can be used to probe heavier pseudoscalar masses than the $\tau\bar\tau$ channel.  Whether the $2\sigma$ excess at high mass from CMS
turns out to be a hint for a signal or just the result of a statistical fluctuation, it seems imperative to probe this 
mass range with a channel 
with comparable or better sensitivity, able to provide complementary information on the MSSM preferred region of parameter space.

\section{Conclusions}
\label{sec:conclusions}

In this article we have studied the associated production of non-standard neutral Higgs bosons with $b$ quarks at the LHC.  Considering the
Higgs  boson decay into $b$ quarks, we have analyzed the reach of the 7 TeV LHC collider.   
After applying a rather simple event selection cuts, a manageable signal-to-background ratio could be obtained, helped
by the enhanced production cross section obtained at large values of $\tan\beta$, which allows for a statistically 
meaningful reach at moderate values of the LHC luminosity. In particular, we find that Higgs bosons with a coupling  
to $b$ quarks of about 0.3 or larger (namely $\tan\beta_{\rm{eff}}^b=30$ or larger),  and with a mass up to 400 GeV could 
be discovered with a luminosity of 30 $\rm{fb}^{-1}$. We expect that the run at 8 TeV will enhance the reach by about 10-15$\%$, although a precise  estimation of the reach depends on many details which are beyond the scope 
of this study. 

We have studied the discovery potential using two different sets of cuts. In general, the $b$ jet produced in association with the Higgs boson tends to be soft, driven by the suppression from the steeply falling PDFs. At the same time, the $p_T$ of the $b$ jets from the Higgs decay is closely correlated to the mass of the Higgs boson.  Therefore, in particular in the low mass region, a somewhat lower threshold on the total jet $p_T$ will enhance 
the discovery reach. At the same time, it could be beneficial to use an asymmetric  $p_T$ selection criteria with the requirement 
that the cut on hardest $b$-jet $p_T$ is correlated with the target Higgs 
mass. This effect should be more prominent for higher Higgs masses.

We have also studied the discovery potential in SUSY-like scenarios. In this case, the corrections to the Yukawa couplings arise at loop level, and there is a correlation between the $b \bar b$ and $\tau \bar \tau$ search channels.  We found that the $3b$ channel can be important in probing 
supersymmetric scenarios in which SUSY-breaking effects can significantly
modify the  couplings of non-standard neutral Higgs bosons to $b$ quarks and 
$\tau$ leptons.
In particular we showed that the $\tau \bar \tau$ channel still has a better reach for 
lower Higgs boson masses, but the $b\bar b$ channel can be used to probe heavier
pseudoscalar masses than the $\tau\bar\tau$ channel.
Furthermore, the $3b$ channel provides an  important probe into the coupling of the Higgs boson to $b$ quarks 
and hence it is complementary to the $\tau\bar\tau$ channel.

\subsubsection*{Acknowledgements}
We would like to thank Johann Alwall, Antonio Boveia, Adam Martin, Pedro Schwaller and Thomas Wright for useful discussions and comments.
Fermilab is operated by Fermi Research Alliance, LLC under Contract No. DE-AC02-07CH11359 with the U.S. Department of Energy. Work at ANL is supported in part by the U.S. Department of Energy~(DOE), Div.~of HEP, Contract DE-AC02-06CH11357. This work was supported in part by the DOE under Task TeV of contract DE-FGO2-96-ER40956. L.T.W. is supported by the NSF under grant PHY-0756966 and the DOE Early Career Award under grant DE-SC0003930. A.M. is supported at University of Oregon by DOE grant number DE-FG02-96ER40969.

\end{document}